\let\ifletter=\iffalse
\newif{\ifarxiv}\arxivtrue

\ifletter
	\ifarxiv
	\documentclass[aps,prl,twocolumn,letterpaper,nofootinbib,superscriptaddress]{revtex4}
		\newlength{\mylength}
		\usepackage[centering,textwidth=\textwidth,textheight=\textheight]{geometry}
	\else
		\documentclass[aps,prl,twocolumn,letterpaper,nofootinbib,superscriptaddress]{revtex4-1}
	\fi
	\usepackage{graphicx,epsf}
	\usepackage{subfigure}
	\usepackage{amsmath,amssymb}
	\usepackage[bookmarks=true,hyperfigures=true]{hyperref}
	\ifarxiv
		\makeatletter
		\def\section{%
		  \@startsection
			{section}%
			{1}%
			{\z@}%
			{0.4cm \@plus1ex \@minus .2ex}%
			{0.1cm}%
			{\centering\normalfont\small\bfseries}% why not centered without \centering?
		}%
		\makeatother
	\fi
\else
	\documentclass[11pt,a4paper]{article}
	\usepackage{graphicx}
	\usepackage{amsmath,amssymb}
	\usepackage[bookmarks=true,hyperfigures=true]{hyperref}
	\usepackage[left=2.8cm,right=2.8cm,top=2.8cm,bottom=2.8cm]{geometry}
\fi

\providecommand{\hypersetup}[1]{}

\providecommand{\texorpdfstring}[2]{#1}
\providecommand{\pdfbookmark}[3][]{}

\hypersetup{plainpages=false}
\hypersetup{pdfpagemode=UseOutlines}
\hypersetup{bookmarksnumbered=true}
\hypersetup{bookmarksopen=true}
\hypersetup{pdfstartview=FitH}
\hypersetup{colorlinks=false}
\hypersetup{citebordercolor={.6 .9 .6}}
\hypersetup{urlbordercolor={.7 .8 1}}
\hypersetup{linkbordercolor={1 .7 .7}}
\hypersetup{pdfborder={0 0 .5}}

\makeatletter
\let\@myabstract\@empty
\let\@keywords\@empty
\let\@subject\@empty
\providecommand{\affiliation}[1]{\gdef\@affiliation{#1}}
\providecommand{\myabstract}[1]{\gdef\@myabstract{#1}}
\providecommand{\keywords}[1]{\gdef\@keywords{#1}}
\providecommand{\subject}[1]{\gdef\@subject{#1}}
\def\thetitle{\@title}
\def\theauthor{\@author}
\def\theaffiliation{\@affiliation}
\def\theabstract{\@myabstract}
\def\thesubject{\@subject}
\def\thedate{\@date}
\def\thekeywords{\@keywords}
\makeatother
\def\fillpdfdata{
\hypersetup{pdftitle={\thetitle}}%
\hypersetup{pdfsubject={\thesubject}}%
\hypersetup{pdfkeywords={\thekeywords}}%
}

\let\oldbfseries=\bfseries
\let\oldmdseries=\mdseries
\let\oldnormalfont=\normalfont
\renewcommand{\bfseries}{\oldbfseries\boldmath}
\renewcommand{\mdseries}{\oldmdseries\unboldmath}
\renewcommand{\normalfont}{\oldnormalfont\unboldmath}

\makeatletter
\newlength{\apb@width}
\newcommand{\autoparbox}[2][c]{\settowidth{\apb@width}{#2}\parbox[#1]{\apb@width}{#2}}
\newcommand{\includegraphicsbox}[2][]{\autoparbox{\includegraphics[#1]{#2}}}
\makeatother

\def\[{\begin{equation}}
\def\]{\end{equation}}
\def\<{\begin{eqnarray}}
\def\>{\end{eqnarray}}

\newcommand{\nn}{\nonumber}

\newcommand{\eps}{\epsilon}
\newcommand{\ar}{i}
\newcommand{\br}{j}
\newcommand{\ccr}{k}

\begin{document}

%%%%%%%%%%%%%%%%%%%%%%%%%%%%%%%%%%%%%%%%%%%%%%%%%%%%%%%%%%%%
%title data
%%%%%%%%%%%%%%%%%%%%%%%%%%%%%%%%%%%%%%%%%%%%%%%%%%%%%%%%%%%%

\title{New Relations for Three-Dimensional Supersymmetric Scattering Amplitudes}

\myabstract{%
We provide evidence for a duality between color and kinematics in three-dimensional supersymmetric Chern--Simons matter theories. We show that
the six-point amplitude in the maximally supersymmetric, ${\cal N}=8$, theory can be arranged so that the kinematic factors satisfy the
fundamental identity of three-algebras. We further show that the four- and six-point ${\cal N}=8$ amplitudes can be ``squared" into the
amplitudes of ${\cal N}=16$ three-dimensional supergravity, thus providing evidence for a hidden three-algebra structure in the dynamics of the
supergravity.}


\noindent
UUITP-06/12
\ifletter

\fi
\hfill
AEI-2012-019

\ifletter

    \author{Till Bargheer}
    \email{till.bargheer@physics.uu.se}
    \affiliation{Department of Physics and Astronomy, Uppsala University, SE-751 08 Uppsala, Sweden}
    \author{Song He}
    \email{songhe@aei.mpg.de}
    \author{Tristan McLoughlin}
    \email{tmclough@aei.mpg.de}
    \affiliation{Max-Planck-Institut f\"ur Gravitationsphysik, Albert-Einstein-Institut, Am M\"uhlenberg 1, D-14476 Potsdam, Germany}

    \hypersetup{pdfauthor={Till Bargheer, Song He, Tristan McLoughlin}}

    \begin{abstract}
    \theabstract
    \end{abstract}

    \pacs{\hspace{1cm}}

    \maketitle

\else

    \author{Till Bargheer,\texorpdfstring{${}^a$}{} Song He,\texorpdfstring{${}^b$}{} and Tristan McLoughlin\texorpdfstring{${}^b$}{}}

    \hypersetup{pdfauthor={\theauthor}}

    \begin{center}
    {\Large\textbf{\mathversion{bold}\thetitle}\par}
    \vspace{1cm}

    \textsc{\theauthor}
    \vspace{5mm}

    \textit{%
    ${}^a$Department of Physics and Astronomy, Uppsala University, SE-751 08 Uppsala, Sweden\\
    ${}^b$Max-Planck-Institut f\"ur Gravitationsphysik, Albert-Einstein-Institut, Am M\"uhlenberg 1, D-14476 Potsdam, Germany}

    \vspace{3mm}
    \verb+till.bargheer@physics.uu.se, {song.he,tmclough}@aei.mpg.de+
    \par\vspace{1cm}

    \textbf{Abstract}\vspace{5mm}

    \begin{minipage}{12.7cm}%\centering
    \theabstract
    \end{minipage}

    \vspace{1cm}

    \end{center}

\fi

\fillpdfdata

%%%%%%%%%%%%%%%%%%%%%%%%%%%%%%%%%%%%%%%%%%%%%%%%%%%%%%%%%%%%
%%%%%%%%%%%%%%%%%%%%%%%%%%%%%%%%%%%%%%%%%%%%%%%%%%%%%%%%%%%%
%%%%%%%%%%%%%%%%%%%%%%%%%%%%%%%%%%%%%%%%%%%%%
\section{Introduction}
%%%%%%%%%%%%%%%%%%%%%%%%%%%%%%%%%%%%%%%%%%%%%
Scattering amplitudes have provided a rich vein of insight into the hidden structures underlying our theories of gauge and gravitational
interactions. One particularly suggestive result is the color-kinematics duality discovered by Bern, Carrasco, and Johansson
(BCJ)~\cite{Bern:2008qj}. At tree-level,  color dressed scattering amplitudes in Yang--Mills (YM) theories can, quite generally, be written as  a sum
over cubic graphs \< \label{eq:colordecomp} {\cal A}_n= g^{n-2} \sum_{\begin{subarray}{l}
        i\in {\rm graphs}
 \end{subarray}} \frac{n_i c_i }{\prod_{\ell_i} p^2_{\ell_i}} ,
\> where the $c_i$'s are color structures made from the usual Lie algebra structure constants, and the $n_i$'s are kinematic factors from which
we have removed products of inverse propagators $p^2_{\ell_i}$ associated to internal lines of the respective cubic diagram. BCJ proposed that
there exists a representation of the amplitude such that for any set of color structures related by a Jacobi identity, there is a corresponding
relation between their numerator factors, i.e. \< \label{eq:Jacobi} c_1 + c_2 + c_3 = 0 \Rightarrow n_1 + n_2 + n_3 = 0 ~. \> This duality
implies non-trivial relations between different  tree-level color-ordered subamplitudes, so-called BCJ relations~\cite{Bern:2008qj}.

Moreover, as is well known, via the Kawai--Lewellen--Tye (KLT) relations~\cite{Kawai:1985xq, Berends:1988zp}, such Yang--Mills amplitudes can be
used to express tree-level scattering in related gravity theories. BCJ~\cite{Bern:2008qj, Bern:2010ue} proposed that it is possible to express
gravity amplitudes in terms of the gauge theory data by simply replacing the color factors by another copy of the kinematic numerators and
summing over the same cubic diagrams.

In this note, we provide evidence for a non-trivial analogue of the
color-kinematics duality in supersymmetric Chern--Simons matter (SCS) theories
and for a corresponding ``double-copy" construction leading to the $ E_{8(8)}$ symmetric, three-dimensional ${\cal N}=16$ supergravity.

%%%%%%%%%%%%%%%%%%%%%%%%%%%%%%%%%%%%%%%%%%%%%
\section{\texorpdfstring{${\cal N}=8$}{N=8} SCS scattering amplitudes}
%%%%%%%%%%%%%%%%%%%%%%%%%%%%%%%%%%%%%%%%%%%%%

The maximally supersymmetric Chern--Simons theory, the
Bagger--Lambert--Gustavsson\ifletter\else\linebreak\fi(BLG) theory, constructed
in~\cite{Bagger:2006sk,Gustavsson:2007vu, Bagger:2007jr}, is the unique
three-dimensional gauge theory with $OSp(8|4)$ superconformal symmetry. The
on-shell, physical states comprise eight scalars, $X^I$, in the $\bf{8}_v$ of $SO(8)$ and eight fermions, $\Psi^{\dot I}$, in the  $\bf{8}_c$.
An important feature of the original construction was the appearance of three-algebras.  Briefly, a three-algebra is a vector space, $T^a$, $a=1,
\dots, N$, with a trilinear product
\begin{eqnarray*}
[T^a, T^b, T^c]=f^{abc}{}_d T^d\,,
\end{eqnarray*}
where the structure constants $f^{abc}{}_d$ satisfy the
fundamental three-algebra identity, \< \label{eq:threealg}
%0 = f^{[abc}{}_g f^{d]efg}
f^{efg}{}_d f^{abc}{}_g=f^{efa}{}_g f^{bcg}{}_d+f^{efb}{}_g f^{cag}{}_d+f^{efc}{}_g f^{abg}{}_d~.
%-f^{abc}{}_g f^{gdeh}+f^{aeh}{}_g f^{gbcd}+f^{adc}{}_g f^{gehb}+f^{abd}{}_g f^{gehc}=0~.
\> Moreover there is a trace form, $h^{ab}={\rm Tr}(T^a T^b)$, which can be used to raise and lower indices. The structure constants with all
indices raised are completely anti-symmetric, $f^{abcd}=f^{[abcd]}$. All
on-shell fields are three-algebra valued fundamental fields, e.g.\
$X^I=\sum_{a=1}^{N}(X^I)^a\,T^a$. The only known finite-dimensional example is where the three-algebra is four-dimensional, while the structure
constants are proportional to the invariant four-index tensor $f^{a_1 a_2a_3 a_4}\propto \eps^{a_1 a_2a_3 a_4}$.

As we are interested in scattering amplitudes, it is convenient to make use
of the spinor-helicity formalism, whereby three-momenta are expressed
as the product of two-component real spinors:%
\footnote{In fact, only particles with positive energy correspond to real spinors. For negative energies,
$\lambda$ is taken to be purely imaginary.}
$p^{\alpha\beta}=\lambda^\alpha\lambda^\beta$ where $\alpha, \beta=1, 2$. The on-shell fields can be
grouped into a single superfield,%
\footnote{Our construction closely parallels the oscillator construction of
the $OSp(8|4)$ algebra~\cite{Gunaydin:1985tc}
and so is only $U(4|2)$ covariant, corresponding to the Jordan decomposition of $OSp(8|4)$ with respect to a $U(1)\in
U(4|2)$. An equivalent on-shell superfield formulation of BLG was constructed in~\cite{Huang:2010rn}.}
by introducing four Gra{\ss}mann parameters $\gamma^i$, $i=1,\dots, 4$. This construction breaks manifest $SO(8)$ R-symmetry by rewriting the
$\bf{8}_v$ scalars as $X^I=\{\bar{X}, X^{[ij]},X\}$ and similarly for the fermions, $\Psi^{\dot A}=\{\psi_i, \bar{\psi}^i\}$, so that the
on-shell superfield is
\<
\Phi_{\rm BLG}&=&\bar{X}+\gamma^\ar { \psi}_\ar+\frac{1}{2}\epsilon_{\ar\br\ccr l} \gamma^\ar\gamma^\br X^{[\ccr l]}\nn\\
& &+\frac{1}{3!}\epsilon_{\ar\br\ccr l}\gamma^\ar\gamma^\br\gamma^\ccr{\bar \psi}^l+\frac{1}{4!}\epsilon_{\ar\br\ccr l}\gamma^\ar\gamma^\br\gamma^\ccr\gamma^l X~.
\label{eqn:superfields}
\>
$OSp(8|4)$ invariant four-point scattering amplitudes,
\<
\label{eq:4ptCS}
{\cal A}_4=\frac{4\pi i}{k}  \frac{\delta^{(3)}(P)\delta^{(8)}(Q)}{\langle 12\rangle\langle 23\rangle\langle 31\rangle}f^{a_1a_2a_3a_4}~,
\>
have previously been constructed~\cite{Huang:2010rn}. In this formula, the delta-functions impose conservation of momenta,
$P^{\alpha\beta}=\sum_{j=1}^4 p^{\alpha\beta}_j$, and supermomenta, $Q^{\alpha i }=\sum_{j=1}^4 \lambda^{\alpha}_j \gamma_j^i$, while the
kinematic invariants are defined as $\langle j k\rangle=\eps_{\alpha\beta}\lambda^\alpha_j\lambda^\beta_k$. The overall form of the amplitude is
fixed by the superconformal symmetries, while the normalization, dependence on the Chern--Simons coupling $k$, and color structure, are fixed by
explicit Feynman diagram calculation of any component amplitude.

Quite generally we can write an $n$-point amplitude in the BLG theory in
the form~\eqref{eq:colordecomp}, but with the $c_i$ corresponding to
three-algebra color structures.%
\footnote{Note that the gauge field is non-dynamical and thus only fundamental matter fields appear as external states.}
The sum is now over diagrams with quartic vertices, and the color structures are found by associating to each
vertex a factor $f^{abcd}$, and to each internal line a metric $h_{ab}$.
For example, Fig.~\ref{fig:quarticdiag} corresponds to
$c_{(123)(456)}:=f^{a_1a_2a_3 b}h_{bc}f^{c a_4a_5a_6 }$.
%%%%%%%%%%%%%%%%%%%%%%%%%%%%%%%%%%%%%%%%%%%%%%%%%
%Figures
\ifletter
	\begin{figure}\centering
	\includegraphics[width=1.5cm, height=1.8cm]{FigColorTree.mps}
	\caption{Quartic diagrams}
	\label{fig:quarticdiag}
	\end{figure}

	\begin{figure}
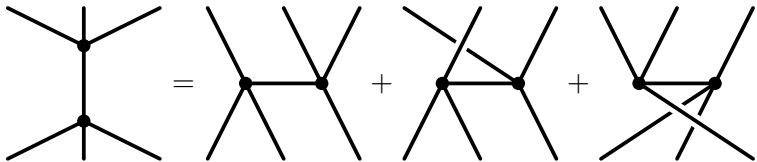
\centering
	$\includegraphicsbox[width=1.5cm, height=1.3cm]{FigColorTree1.mps}=\includegraphicsbox[width=1.5cm,height=1.3cm]{FigColorTree2.mps}+
	\includegraphicsbox[width=1.5cm,height=1.3cm]{FigColorTree3.mps}+
	\includegraphicsbox[width=1.5cm,height=1.3cm]{FigColorTree4.mps}$
	\caption{Graphical expression of the fundamental identity.}
	\label{fig:funiden}
	\end{figure}
\else
	\begin{figure}\centering
	\includegraphics{FigColorTree.mps}
	\caption{Six point quartic diagram.}
	\label{fig:quarticdiag}
	\end{figure}

	\begin{figure}\centering
	$\includegraphicsbox{FigColorTree1.mps}=\includegraphicsbox{FigColorTree2.mps}+
	\includegraphicsbox{FigColorTree3.mps}+
	\includegraphicsbox{FigColorTree4.mps}$
	\caption{Graphical expression of the fundamental identity.}
	\label{fig:funiden}
	\end{figure}
\fi
%%%%%%%%%%%%%%%%%%%%%%%%%%%%%%%%%%%%%%%%%%%%%%%%
A key feature is that due to the fundamental identities \eqref{eq:threealg} not all of the color structures are independent. Namely, given
$c_s=\dots f^{efg}{}_d f^{abc}{}_g\dots $, $c_t=\dots f^{efa}{}_g f^{bcg}{}_d\dots $, $c_u=\dots f^{efb}{}_g f^{cag}{}_d\dots $ and $c_v=\dots
f^{efc}{}_g f^{abg}{}_d\dots$, where the ``$\dots$" denote factors common to all diagrams, then $c_s=c_t+c_u+c_v$. Our first proposal is that
corresponding numerators $n_s$, $n_t$, $n_u$ and $n_v$ can always be found such that (see Fig.~\ref{fig:funiden})
\< \label{eq:funiden}
c_s=c_t+c_u+c_v \Rightarrow n_s=n_t+n_u+n_v~.
\>

We do not have a general proof for these relations, instead we will provide evidence for their existence by considering the first non-trivial
case, i.e.\ six points.

At six points, all color structures consist of the contractions of two tensors as in Fig.~\ref{fig:quarticdiag}.
Accounting for the anti-symmetry of $f^{abcd}$, there are ten distinct color structures $c_i$, labeled by partitions of the six color labels
into groups of
three, e.g.\ $c_{1}=c_{(123)(456)}$.%
\footnote{To be explicit we will also choose $c_2=c_{(156)(234)}$, $c_3=c_{(612)(345)}$,
$c_4=c_{(125)(436)}$, $c_5=c_{(136)(245)}$, $c_6=c_{(145)(236)}$, $c_7=c_{(124)(356)}$, $c_8=c_{(143)(256)}$, $c_9=c_{(146)(235)}$,
$c_{10}=c_{(135)(246)}$. We use the same notation for labeling the numerators.}
At six points there are five independent three-algebra relations between different color structures. Our claim is that there is a choice of
numerators such that they satisfy the same three-algebra fundamental identities, however the numerators are not uniquely defined and finding
explicit forms is not straightforward. Instead, we will show the existence of such numerators, and give a recipe for calculating them,  by
considering the color-ordered subamplitudes of ${\cal N}=6$ Aharony--Bergman--Jafferis--Maldacena (ABJM) theory \cite{Aharony:2008ug}. 

%%%%%%%%%%%%%%%%%%%%%%%%%%%%%%%%%%%%%%%%%%%%
\section{New relations for color-ordered subamplitudes}
%%%%%%%%%%%%%%%%%%%%%%%%%%%%%%%%%%%%%%%%%%%%

As is well known, the BLG theory can be rewritten~\cite{VanRaamsdonk:2008ft} as a special case ($N=2$) of the $SU(N)\times SU(N)$ ${\cal
N}=6$ Chern--Simons theories with bi-fundamental matter, that is ABJM-theories. 
The ABJM on-shell fields can be grouped
into two superfields, $\hat \Phi^{\scriptscriptstyle A}_{\scriptscriptstyle
\bar A}$, transforming as $(N,\bar N)$, and $\hat {\bar
\Phi}^{\scriptscriptstyle\bar B}_{\scriptscriptstyle B}$, transforming as  $(\bar N, N)$~\cite{Bargheer:2010hn}.  This formalism is manifestly
$U(3)$ symmetric, making use of three Gra{\ss}mann parameters $\gamma^{\hat i}$, $\hat i=1,2,3$. For $N=2$ the conjugate representations are
equivalent and the two superfields can be combined: $\Phi_{\rm BLG}={\hat \Phi}+\gamma^4 \hat{\bar \Phi}$.

Scattering amplitudes in BLG theory can be found from those of ABJM by identifying the appropriate fields and color structures. ABJM scattering
amplitudes can however be decomposed into color-ordered subamplitudes. Each color-ordered subamplitude will contribute to several kinematical
coefficients of the BLG color structures $c_i$. We claim that every color-ordered ABJM subamplitude can be written as a certain combination of
numerators $n_i$ with propagators, in such a way that the corresponding BLG amplitudes take the form~\eqref{eq:colordecomp}, with the numerators
satisfying the three-algebra identities \eqref{eq:funiden}. This implies non-trivial relations among the color-ordered ABJM subamplitudes, and
thus is a slightly stronger claim than the proposition that the BLG amplitudes decompose as \eqref{eq:colordecomp} with \eqref{eq:funiden}
satisfied. In the following, we provide evidence for this proposal by examining the six-point amplitudes.

Four-point amplitudes in ABJM%
\footnote{Actually of a one parameter family of mass deformed theories.}
were considered in~\cite{Agarwal:2008pu}, the six-point color-ordered subamplitude for ABJM were first calculated in~\cite{Bargheer:2010hn}, see
also~\cite{Gang:2010gy}.
As a representative component amplitude, we consider the six-point amplitude involving a single flavor of complex scalar $\phi(p)^{A}_{\bar A}$
and its conjugate $\bar \phi(p)^{\bar B}_{ B}$,
\<
\hat A_{6\phi}=
A(1,2,3,4,5,6)\,\delta^{{\scriptscriptstyle \bar B_2}}_{\scriptscriptstyle\bar A_1} \delta^{\scriptscriptstyle A_2}_{\scriptscriptstyle B_2} \delta^{\scriptscriptstyle\bar B_4}_{\scriptscriptstyle\bar A_3}  \delta^{\scriptscriptstyle A_5}_{ B_4}\delta^{\scriptscriptstyle \bar B_6}_{\scriptscriptstyle \bar A_5}  \delta^{\scriptscriptstyle A_1}_{ \scriptscriptstyle B_6}+\dots~,
\label{eq:abjm6phi}
\>
with the ellipses denoting other color orderings.

At six points, we propose that the color-ordered ABJM subamplitudes take
the form

\begin{eqnarray*}
A(i,j,k,p,q,r)=\frac{n_{(ijk)(pqr)}}{p_{ijk}^2}+\frac{n_{(qri)(jkp)}}{p_{qri}^2}+\frac{n_{(rij)(kpq)} }{p_{kpq}^2}~.
\end{eqnarray*}
There are six independent subamplitudes, all others are related to those by cyclic double-shifts and by inversions, e.g.\ifletter $A(3,4,5,6,1,2)=A(1,2,3,4,5,6)$,
$A(1,2,5,6,3,4)=A(1,4,3,6,5,2)$. \else
\begin{align*}
A(3,4,5,6,1,2)&=A(1,2,3,4,5,6)~,\nn\\
A(1,2,5,6,3,4)&=A(1,4,3,6,5,2)~.\nn
\end{align*}
\fi
If, as we claim, it is possible to choose numerators satisfying \eqref{eq:funiden},
we can solve for five of the numerators, for example by setting
\begin{eqnarray*}
n_9=-n_1+n_{10}+n_4~, & &~~~n_8=-n_2+n_3+n_4\\
n_7=n_1-n_3-n_4~, & &~~~n_6=n_{10}-n_2+n_4\\
&&\kern-55pt n_5 =-n_1+n_{10}+n_3+n_4~.
\end{eqnarray*}
We can now solve for four further numerators $n_2$, $n_3$,  $n_4$ and $n_{10}$ in terms
of known expressions~\cite{Bargheer:2010hn, Gang:2010gy} for $A(1,2,3,4,5,6)$, 
 $A(1,2,3,6,5,4)$, $A(1,2,5,4,3,6)$, and 
 $A(1,4,3,6,5,2)$. We thus
derive identities for $A(1,4,5,2,3,6)$ and $A(1,4,3,2,5,6)$ in terms of
these subamplitudes and the undetermined numerator $n_1$. The
expressions for the numerators and correspondingly the identities are rather complicated, however it is straightforward to numerically check, by
choosing explicit numerical values for external momenta, that they are in fact satisfied. Importantly, the undetermined kinematical factor,
$n_1$, does not appear in any of these relations and so corresponds to a
generalized gauge freedom analogous to that found in the YM case~\cite{Bern:2008qj}.

%%%%%%%%%%%%%%%%%%%%%%%%%%%%%%%%%%%%%%%%%%%%
\section{\texorpdfstring{$E_{8(8)}$}{E8(8)} supergravity theory}
%%%%%%%%%%%%%%%%%%%%%%%%%%%%%%%%%%%%%%%%%%%%

The three-dimensional  ${\cal N}=16$ supergravity with $E_{8(8)}$ symmetry ($E_8$-theory), originally constructed by Marcus and
Schwarz~\cite{Marcus:1983hb}, consists of 128 scalar bosons and 128 fermions which are in inequivalent real spinor representations of $SO(16)$,
the maximal compact subgroup of $E_{8(8)}$. An immediate consequence of this, as explained in~\cite{Marcus:1983hb}, is that non-trivial
scattering amplitudes must have an even number of external particles, as products of odd numbers of spinors cannot form a singlet. Consequently
the S-matrix is naively different than the dimensional reduction of the four-dimensional ${\cal N}=8$ supergravity with $E_{7(7)}$ symmetry
($E_7$-theory). However, as is long known, e.g.~\cite{Breitenlohner:1987dg}, on-shell the two theories are related by performing a duality
transformation, after dimensional reduction, of all the vector fields into scalars,  which then combine with the scalars from dimensional
reduction, including those originally in the $E_{7(7)}/SU(8)$ coset of the ${\cal N}=8$ supergravity, to become those of the $E_{8(8)}/SO(16)$
coset.

The $E_{8(8)}$ algebra comprises 120 compact $SO(16)$ generators $X^{IJ}$, $I,J=1,\dots,16$, and 128 non-compact generators $Y^A$, $A=1,\dots,
128$.

It is convenient to fix the unitary-gauge, whereby a generic group element is written as $g=e^{\varphi^A Y^A}$ with  $\varphi^A$  the physical
scalars. The  $E_{8(8)}/SO(16)$-coset action is constructed from the algebra-valued current $P_\mu =\frac{1}{2}\left(e^{-\varphi}\partial
e^{\varphi}-e^{\varphi}\partial e^{-\varphi}\right)$. The bosonic action is~\cite{Marcus:1983hb},
\begin{eqnarray}
{\cal L}_{\rm bos}=\frac{1}{4\kappa^2} \sqrt{-g} R-\frac{1}{4\kappa^2} \sqrt{-g} g^{\mu\nu} P_\mu^A P_\nu^A,
\end{eqnarray}
where the first term is the usual gravity action. Using this action (the fermionic terms are also known), with appropriate  gauge fixing, one can straightforwardly
calculate scattering amplitudes using Feynman diagrams. At four-points such amplitudes for four scalars receive contributions from graviton
exchange and from contact interactions that arise upon expanding the coset term to quartic order in fields, ${\cal L}_{\varphi^4}\sim (\varphi
\Gamma^{IJ}\partial_\mu \varphi)( \varphi \Gamma^{IJ}\partial_\mu \varphi)$. In the simplest case we can consider the scattering of four scalars
all carrying the same coset index, e.g.\ all fields being  $\varphi^1$, in which case there is no contribution from contact terms. Combining all
graviton exchange diagrams we find
\< M_4=\frac{i \kappa^2}{4}\left( \frac{s^2+u^2}{ t}+\frac{t^2+u^2}{ s}+\frac{s^2+t^2}{u}\right)~. \>
It is not difficult to calculate other component amplitudes, however we can make use of the supersymmetry to determine the full four-point
superamplitude.

For the $E_8$-theory we can define an on-shell superfield by using eight
Gra{\ss}mann parameters $\eta^I$, $I=1,\dots, 8$ which breaks the $SO(16)$
R-symmetry to $U(8)$. Splitting the 128 scalars $\varphi^A$ into the fields $\{\xi, {\bar \xi}, \xi_{IJ},{\bar \xi}^{IJ},\xi_{IJKL}\}$ with, for example
$\xi=\tfrac{1}{2}(\varphi^1+i \varphi^2)$, and similarly for the fermionic
fields, we can write the superfield%
\footnote{This superfield is very
similar to that of the $E_7$-theory  and indeed making the formal identification
$\xi=h$, $\bar \xi=\bar h$, $\xi_{IJ}=B_{IJ}$, $\bar \xi^{IJ}={\bar B}^{IJ}$, $ \xi_{IJKL}=D_{IJKL}$ to the fields
of the  $E_7$-theory this becomes more apparent.}
\< \Xi= \xi+\eta^I  \psi_I+\frac{1}{2}\eta^I\eta^J \xi_{IJ}+\dots+\frac{1}{8!}\eta^8{\bar \xi}~. \>

By using super-Poincar\'e symmetry and matching to the component amplitude, the four-point superamplitude is
\< \label{eq:E8fourpt} {\cal M}_4 =\frac{i \kappa^2}{4}\frac{\delta^{(16)}(Q)\delta^{(3)}(P)}{(\langle 12\rangle\langle 23\rangle\langle
31\rangle)^2}~. \>
Here, the 16-dimensional fermionic delta-function is given by the product of two eight-dimensional fermionic delta-functions, $\delta^{(16)}(Q)\sim
\delta^{(8)}(Q^1)\delta^{(8)}( Q^2)$, such as appeared in \eqref{eq:4ptCS}. Stripping off the overall normalization and momentum delta-function
we see that this is the ``square" of \eqref{eq:4ptCS}. This then suggests an analogue of the KLT relation~\cite{Kawai:1985xq,Berends:1988zp}
between ${\cal N}=4$ supersymmetric Yang--Mills (SYM) and the $E_7$ supergravity theory to one between ${\cal N}=8$ BLG and the $E_8$-theory. As
zeroth order checks, we note that the spectra of the $E_8$-theory and that of the BLG theory squared match; furthermore in both cases all
non-trivial amplitudes have even numbers of legs. Of course the direct dimensional reduction of  ${\cal N}=4$ SYM and the $E_7$-theory amplitudes to three
dimensions are related by the usual KLT relations, and for fields which are unchanged by the duality transformation, in particular the scalars
originating in the $E_{7(7)}/SU(8)$ coset, the three-dimensional scattering amplitudes are just those of the four-dimensional theory evaluated
on three-dimensional kinematics. However, after the duality transformation to the $E_8$-theory, this ceases to be the case for all amplitudes;
as a simple example there is no $E_8$-theory three-point amplitude corresponding to the square of the three-dimensional SYM three-point
amplitude. As ${\cal N}=8$ BLG theory can be found from supersymmetric three-dimensional Yang--Mills~\cite{Mukhi:2008ux} via a
``Higgsing"-procedure reminiscent of the duality transformation, it is perhaps not surprising that it should be thus related to the $E_8$
supergravity theory.

%%%%%%%%%%%%%%%%%%%%%%%%%%%%%%%%%%%%%%%%%%%%
\section{Three-dimensional gravity as the square of Chern--Simons}
%%%%%%%%%%%%%%%%%%%%%%%%%%%%%%%%%%%%%%%%%%%%

Given the suggestion that the BLG amplitudes can be written in terms of numerators
satisfying the three-algebra color-kinematics duality, it is natural to ask if the gravity
theory amplitudes can be written as  a ``double-copy" as in~\cite{Bern:2008qj},
\<
\label{eq:gravsquare}
M_n=i \left(\frac{\kappa}{2}\right)^{n-2}\sum_{\begin{subarray}{l}
        i
      \end{subarray}} \frac{n_i n_i }{\prod_{\ell_i} p^2_{\ell_i}}\,,
\> where the $n_i$'s are the numerators appearing in the BLG amplitude \eqref{eq:colordecomp} and the sum is over the same $n$-point quartic
diagrams. This relation obviously holds at four points for the superamplitudes, and at six points we can perform an explicit check by making use
of the numerators calculated from the six-point color-ordered ABJM subamplitudes for specific components. For example, the pure scalar ABJM
amplitude $\hat A_{6\phi}$ \eqref{eq:abjm6phi} can be used to calculate the numerators for the $A_6(X_1 {\bar X}_2X_3 {\bar X}_4X_5 {\bar X}_6)$
in the BLG theory, which potentially squares into the $M_6(\xi_1 {\bar \xi}_2 \xi_3 {\bar \xi}_4\xi_5 {\bar \xi}_6)$ gravity amplitude. That this is indeed
the case can in principle be shown by comparing with the result of a direct Feynman diagram calculation. Equivalently, but significantly more
efficiently, one can take this complex scalar to have originated in the $E_{7(7)}/SU(8)$ coset, so that the squared amplitude can be compared
with the dimensional reduction of the six-scalar $E_7$ supergravity amplitude.
The latter can be found from a scalar component of ${\cal N}=4$ SYM Next-to-Maximally-Helicity-Violated (NMHV) amplitude, conveniently written
as a sum of the so-called R-invariants, multiplied by an MHV pre-factor~\cite{Drummond:2008vq}, \< {\cal A}^{\rm NMHV}_6={\cal A}^{\rm MHV}_6\times
\sum_{3\leq s+1 < t \leq 5}R_{6st}\nn \> and by making use of the KLT relations~\cite{Kawai:1985xq,Berends:1988zp}. It is then straightforward
to check, again by choosing a range of numerical values for external momenta, that the resulting pure scalar amplitude in fact agrees with the
squared BLG amplitude \eqref{eq:gravsquare}.

For higher-point amplitudes it would be possible to prove, along the lines of~\cite{Bern:2010yg}, that \eqref{eq:gravsquare} holds if there were
Britto--Cachazo--Feng--Witten (BCFW) recursion relations~\cite{Britto:2004ap, Britto:2005fq} for the $E_8$-theory. Recursion relations for ABJM
theories, and thus BLG theories, have been proven in~\cite{Gang:2010gy}. The key step is proving that the superamplitude falls off sufficiently
fast for large deformations of the momenta under a complex non-linear shift: $ \hat {\cal A}(\{\lambda_1(z), \lambda_l(z)\}) \sim{\cal
O}\left(1/z\right) $ as $z\rightarrow \infty$ with $\lambda_1(z)=\tfrac{z+z^{-1}}{2} \lambda_1-\tfrac{z-z^{-1}}{2i } \lambda_l$,
$\lambda_l(z)=\tfrac{z-z^{-1}}{2i} \lambda_1+\tfrac{z+z^{-1}}{2 } \lambda_l$ and similar shifts for the Gra{\ss}mann parameters. The proof of a
sufficient fall-off for $E_8$ superamplitudes does not currently exist. However, it is possible to naively apply the method
of~\cite{Gang:2010gy} and use the four-point amplitude \eqref{eq:E8fourpt} to construct a candidate six-point superamplitude in $E_8$
supergravity. We find that the relevant scalar component, 
$M_6(\xi_1 {\bar \xi}_2 \xi_3 {\bar \xi}_4\xi_5 {\bar \xi}_6)$, of this superamplitude agrees with
the amplitude calculated by squaring the numerators \eqref{eq:gravsquare}. This shows that at least to six points the BCFW recursion relations
of~\cite{Gang:2010gy} hold for the $E_8$-theory.

%%%%%%%%%%%%%%%%%%%%%%%%%%%%%%
\section{Outlook}

In order to confirm the proposed ``double-copy" relations for the $E_8$-theory, it would be very useful to prove in general the BCFW relations
for the three-dimensional supergravity. Relatedly, numerator identities for YM and squaring relations for gravity have been conjectured to extend to all-loop diagrams~\cite{Bern:2010ue}, and it would be interesting to check whether similar relations hold for the
three-dimensional Chern--Simons and gravity theories beyond tree-level. If this does indeed hold it would demonstrate the existence of a hidden
three-algebra structure in three-dimensional gravity. This is interesting as a non-trivial model for similar structures in four dimensional
gravity, particularly with regard to knotty issues of quantum gravity, and as an important intermediary step to two-dimensions where gravity is
known to posses infinite dimensional symmetries  \cite{Geroch:1972yt, Julia:1980gr}.

%%%%%%%%%%%%%%%%%%%%%%%%%%%%%%%%%%%%%%%%%%%%%%%%%
% Acknowledgments
%%%%%%%%%%%%%%%%%%%%%%%%%%%%%%%%%%%%%%%%%%%%%%%%%
\ifletter
	\medskip
\else
	\subsection*{Acknowledgements}
\fi

We thank N. Beisert, G. Bossard, F. Loebbert, 
R. Monteiro, H. Nicolai, D. O'Connell,  and ~ R. Roiban
for helpful discussions. We would further
like to thank the organizers of the Nordita
program on Exact Results in Gauge-String 
Dualities, where parts of this work were presented,
for their hospitality. Finally, TMcL would like to thank
CPTh, Ecole Polytechnique for their hospitality 
during the completion of this work.

%%%%%%%%%%%%%%%%%%%%%%%%%%%%%%
%Bibiliography
%%%%%%%%%%%%%%%%%%%%%%%%%%%%%%
\ifletter
	\bibliographystyle{apsrev}
\else
	\bibliographystyle{nb}
\fi
\bibliography{CS}

\end{document}